\documentclass[journal]{IEEEtran}
\usepackage{hyphenat}

\usepackage{wrapfig}
\usepackage{tikz}
\usepackage{pgfplots}
\usepackage{graphicx}
\usepackage[utf8]{inputenc}
\usepackage{tikz}
\usepackage{tcolorbox}
\tcbuselibrary{most}
\usepackage{amsmath,amssymb}
\usepackage{xcolor}
\usepackage{tikz}
\usetikzlibrary{arrows.meta, positioning}
\usepackage[utf8]{inputenc} 
\usepackage[T1]{fontenc}    
\usepackage{hyperref}       
\usepackage{url}            
\usepackage{booktabs}       
\usepackage{amsfonts}       
\usepackage{nicefrac}       
\usepackage{microtype}      
\usepackage{xcolor}         
\usepackage{wrapfig}
\usepackage{tikz}
\usepackage{pgfplots}
\title{Agents Should Replace Narrow Predictive AI as the Orchestrator in 6G AI-RAN}

\author{%
  Pranshav Gajjar$^1$, and Vijay K Shah$^1$\\
  $^1$NextG Wireless Lab,\\
  North Carolina State University\\
}

\begin{document}

\maketitle

\begin{abstract}
This position paper argues that to achieve Level 5 autonomous 6G networks, the next generation of Artificial Intelligence in Radio Access Networks (AI-RAN) should transition away from fragmented, narrow predictive models and instead adopt multimodal Large Language Models (LLMs) as central reasoning agents. Current AI-RAN architectures rely on disjointed Deep Neural Networks (DNNs) and Deep Reinforcement Learning (DRL) agents that operate in isolated domains. These narrow models suffer from siloed knowledge, severe brittleness to out-of-distribution dynamics, and a fundamental inability to bridge the \textit{intent gap} the semantic disconnect between high-level, unstructured operator directives and rigid numerical network configurations. We propose elevating LLMs, or domain-adapted Large Telecom Models (LTMs), to act as the \textit{cognitive operating system} situated within the RAN Intelligent Controller (RIC), the control and orchestration layer of AI-RAN. In this architecture, LLMs do not replace narrow models but orchestrate them as executable subroutines, dynamically translating human intent into concrete policies and utilizing Retrieval-Augmented Generation (RAG) to autonomously diagnose complex, multi-vendor network anomalies. To make this architectural shift a reality, we call upon the machine learning community to prioritize critical foundational research tailored to the strict constraints of telecommunications, specifically focusing on continuous alignment via network-driven feedback (RLNF), extreme sub-8-bit edge quantization, neuro-symbolic verification to curb hallucinations, and securing orchestration frameworks against adversarial prompt injections.

\end{abstract}

\section{Introduction}

The evolution of cellular networks toward the 6G era represents a paradigm shift from conventional, rule-based infrastructure to natively intelligent systems, broadly conceptualized as the Artificial Intelligence-Radio Access Network (AI-RAN) \cite{khan2023ai}. Within this evolving ecosystem, the O-RAN (O-RAN) architecture has democratized network control by introducing the RAN Intelligent Controller (RIC), providing standardized interfaces to host third-party machine learning applications \cite{alliance2023ran}. To date, the telecommunications and machine learning communities have heavily invested in applying AI to optimize specific layers of the protocol stack. Traditional AI applied to telecom has become synonymous with localized optimization: utilizing Deep Reinforcement Learning (DRL) for MAC layer resource block scheduling \cite{villegas2025drl}, deploying localized Deep Neural Networks (DNNs) for channel estimation  \cite{ferrand2020dnn, balevi2020massive}, and employing time series forecasting for predictive user mobility  \cite{xu2016big}.

However, while these narrow, task-specific predictive models excel at isolated mathematical optimization, they fall critically short when confronted with the holistic realities of network orchestration and management. The management plane of modern cellular architectures is exceptionally complex, highly dynamic, and heavily reliant on unstructured data  \cite{zhang2025survey, petrovic2026llm}. Network operators are routinely inundated with cascading, cryptically formatted syslog messages, multi-vendor interoperability anomalies, and high-level business objectives that should be manually translated into rigid numerical configurations  \cite{hossain2026ai}. 

Narrow ML models are inherently blind to this broader operational context. A DRL agent optimized strictly for spectral efficiency cannot parse 3GPP technical specifications, nor can it deduce that an anomalous latency spike is the result of an unmodeled hardware degradation documented in a vendor's unstructured service log  \cite{chen2023tele}. This semantic gap between high-level operational intent and low-level, multi-variable network execution forms the primary bottleneck preventing the telecommunications industry from achieving Level 5 autonomous networks  \cite{wei2025large}. To overcome this, the architecture of AI-RAN should evolve from relying on fragmented, narrow predictive models to utilizing centralized, multimodal reasoning agents.

\textbf{We argue that the machine learning community should elevate Large Language Models (LLMs) from peripheral IT support tools to the core cognitive engines of the AI-RAN control plane, replacing disjointed predictive models as the primary orchestrators of 6G networks.}

By deploying multimodal LLMs or domain-adapted Large Telecom Models (LTMs) as autonomous reasoning agents within the RIC, we can fundamentally transform network orchestration  \cite{wei2025large, petrovic2026llm}. Unlike narrow models, LLMs possess the emergent ability to parse unstructured natural language, reason over complex technical documentation via Retrieval Augmented Generation (RAG)  \cite{gajjar2025oran}, and dynamically generate executable code and configuration scripts  \cite{gajjar2025oransight, wei2025large}. In our proposed architecture, the LLM acts as the central \textit{cognitive operating system}: it deconstructs human intent, contextualizes network anomalies across multiple domains, and actively orchestrates lower-level, narrow predictive models as localized \textit{tools} to execute specific real-time tasks.

In this position paper, we outline the severe limitations of the current narrow AI-RAN paradigm and establish the necessity of an agentic, LLM-driven architecture. We explore how integrating LLM agents into the management plane resolves the intent translation gap and enables zero-touch fault diagnosis. Finally, we address valid counterarguments regarding inference latency and mission-critical hallucination  \cite{huang2025survey}, explicitly calling upon the machine learning community to prioritize foundational research in extreme edge quantization, neuro-symbolic verification, and continuous network-aligned reinforcement learning to realize this necessary architectural shift.

\section{Why AI-RAN?}

For decades, the telecommunications industry relied on monolithic, proprietary hardware solutions, rendering the Radio Access Network (RAN) a \textit{black box} inaccessible to external algorithmic innovation. The introduction of the O-RAN architecture dismantled this barrier by explicitly disaggregating software from baseband hardware and standardizing the interfaces between network components \cite{tripathi2025fundamentals}. For the machine learning community, O-RAN is nothing short of revolutionary: it exposes high fidelity, real-time network telemetry and provides standardized execution environments, effectively turning global cellular infrastructure into a massive, programmable sandbox for ML researchers to deploy and test their models.

At the heart of this programmable architecture is the RAN Intelligent Controller (RIC), which serves as the designated operating system for hosting third-party network optimization algorithms  \cite{polese2023understanding, alliance2023ran}. To manage the vastly different latency requirements of telecommunications, the RIC is structurally divided into two distinct domains:

\begin{itemize}
    \item \textbf{The Non Real Time (Non RT) RIC:} Situated centrally within the service management and orchestration framework, this component governs processes operating on timescales greater than one second. It is the natural home for compute-heavy ML tasks, such as model training, large-scale historical data analytics, and the execution of \textit{rApps}, which dictate overarching network policies and long-term optimization strategies.
    \item \textbf{The Near Real Time (Near RT) RIC:} Deployed closer to the physical network edge, this controller manages closed-loop control functions operating strictly between 10 milliseconds and one second. It hosts \textit{xApps}, which are lightweight, fast-acting microservices responsible for rapid ML inference tasks like dynamic traffic steering, active beamforming optimization, and real-time interference mitigation.
\end{itemize}

While O-RAN provides the necessary open plumbing, simply having open interfaces is no longer enough. The sheer scale and complexity of upcoming 6G networks, characterized by ultra-dense multi-vendor deployments and highly dynamic network slicing, far exceed the capabilities of traditional rule-based engineering and human-driven heuristics. This operational reality necessitates the transition to the AI-RAN paradigm \cite{khan2023ai}. AI-RAN goes beyond merely \textit{patching} machine learning onto existing infrastructure; it envisions a network where AI is the native, foundational fabric. By natively integrating deep learning and intelligent agents directly into the RICs and underlying baseband units, AI-RAN enables the infrastructure to proactively predict traffic fluctuations, self-optimize radio resources, and dynamically adapt to novel environmental variables. AI-RAN is the definitive future of telecommunications because it provides the only scalable architectural paradigm capable of handling the exponential complexity of next-generation networks while driving toward fully autonomous, zero-touch operations.

Because these distinct concepts, O-RAN and AI-RAN, are frequently conflated in emerging literature, it is crucial to explicitly delineate their synergistic relationship \cite{polese2026beyond}. O-RAN fundamentally provides the structural prerequisite, the disaggregated skeleton, and a standardized nervous system that makes algorithmic intervention possible. Conversely, AI-RAN serves as the cognitive brain built directly atop this foundation. Rather than treating the terms as interchangeable, one must view O-RAN as the architectural enabler, and AI-RAN as the resulting evolutionary shift that maximizes its programmable potential.

\section{The Limitations of Narrow AI}
\begin{figure}
    \centering
    \includegraphics[width=\linewidth]{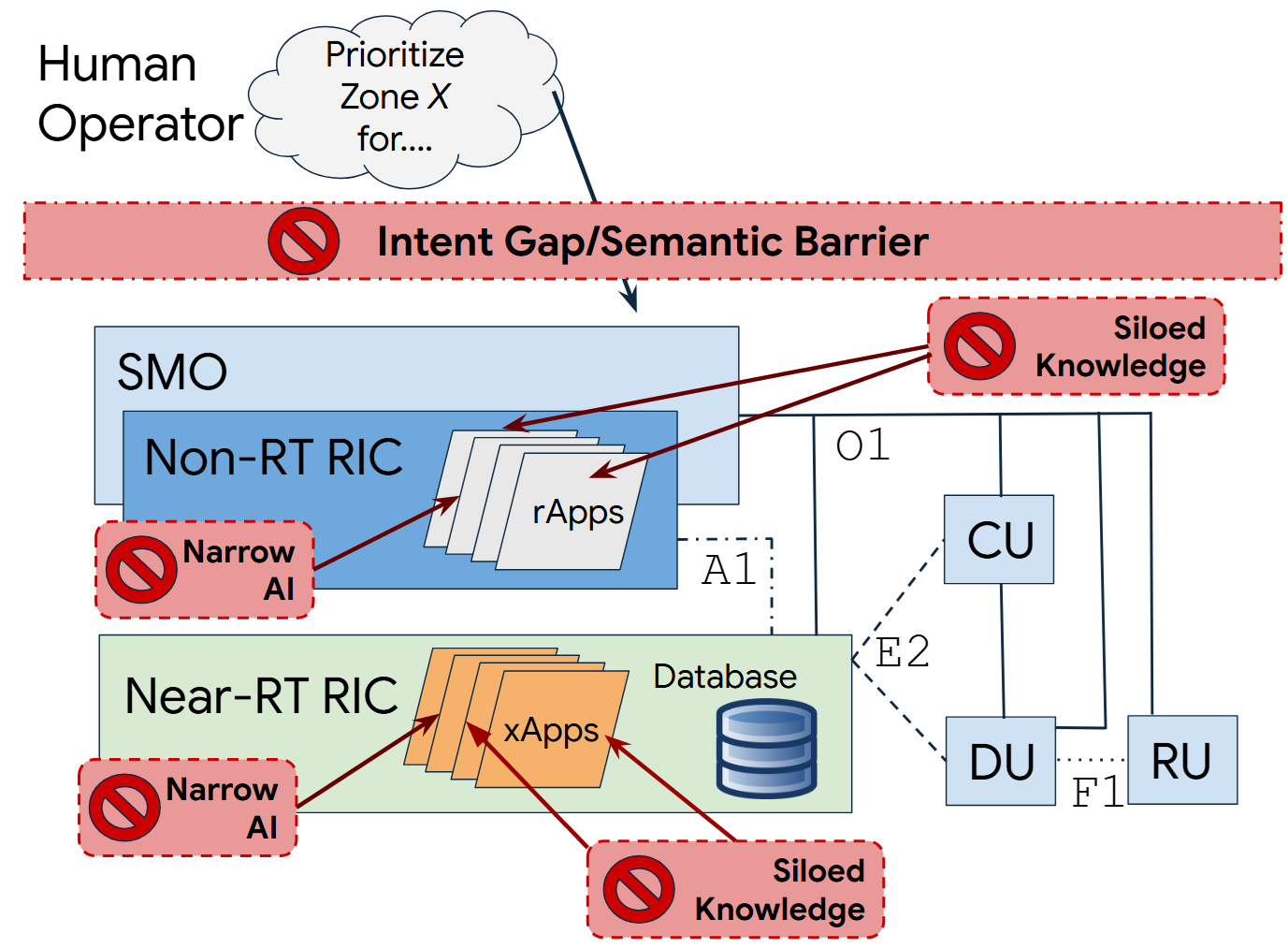}
    \caption{Current fragmented O-RAN architecture. Narrow ML models operate in isolated domains across the O-RAN stack without a centralized cognitive framework to orchestrate them or interpret high-level operator intent. Here, A1, O1, E2, and F1 represent the different interfaces \cite{polese2023understanding} between the two RICs and different O-RAN components.}
    \label{fig:narrow_ai_architecture}
\end{figure}The current trajectory of machine learning integration within telecommunications has inadvertently led to an architecture of fragmented, hyper-specialized intelligence. While the O-RAN alliance has successfully standardized interfaces to decouple software from hardware, the resulting AI-RAN/O-RAN deployments rely on embedding purpose-built, narrow predictive models across the protocol stack  \cite{polese2023understanding}. As illustrated in Figure \ref{fig:narrow_ai_architecture}, this approach peppers the network with isolated mathematical optimizers that lack any centralized cognitive oversight or semantic awareness.


To understand why this paradigm is insufficient for achieving Level 5 autonomy, we should examine the fundamental bottlenecks of narrow predictive models in complex operational environments:

\subsection{Siloed Knowledge and the Multi-Objective Trap}
Narrow models are intrinsically confined to their localized observation spaces and single-objective reward structures  \cite{navidan2026toward}. Within the O-RAN architecture, this isolation inevitably manifests as direct operational conflicts between disparate control applications. For instance, an rApp in the Non-Real-Time RIC may enforce overarching network policies, such as aggressive energy-conservation strategies \cite{del2025pacifista, giannopoulos2025comix}. Simultaneously, a lightweight xApp hosted in the Near-Real-Time RIC might act to maximize localized spectral efficiency by aggressively scheduling resource blocks or executing dynamic traffic steering. Because these fragmented models operate without holistic oversight or semantic data sharing, they inevitably optimize at cross-purposes. The mathematical success of the xApp's performance optimization directly undermines the rApp's power-saving directives, which can induce sub-optimal performance, unexpected latency spikes, or severe instability across different layers of the network \cite{salmi2025ai, del2025pacifista, kwon2026open}. Without a central, cognitively aware orchestrator to actively manage these conflicts and adjudicate multi-objective trade-offs, the network state will oscillate unpredictably as isolated models continuously compete against one another

\subsection{Brittleness to Out-of-Distribution Dynamics}
The operational reality of a cellular network is profoundly non-stationary, constantly shaped by external physical environments, hardware degradation, and unpredictable user behaviors  \cite{cui2026review}. Narrow predictive AI predominantly trained on historical datasets or highly constrained simulated environments exhibits severe brittleness when confronted with these inevitable distribution shifts \cite{liu2023leaf} often requiring retraining to tackle Model drift \cite{manias2023model}, and concept drift \cite{liu2023leaf}. Whether facing zero-day hardware anomalies, atypical human mobility patterns during unforeseen large-scale events, or novel multi-vendor interoperability failures, traditional Deep Neural Networks (DNNs) lack the capacity for zero-shot generalization. When a narrow model encounters an unrepresented state, it neither gracefully degrades nor can it seek diagnostic context; instead, it confidently executes erratic policies or produces catastrophic prediction errors.

\subsection{The Intent Gap and Semantic Disconnect}
Perhaps the most critical limitation to true network autonomy is the insurmountable semantic barrier between human operational objectives and the narrow machine-learning execution. Network management is fundamentally driven by high-level business goals, Service Level Agreements (SLAs), and compliance mandates, all of which are articulated in natural language and unstructured concepts  \cite{jiang2026agentic}. Currently, network operators should manually translate these abstract goals into rigid, numerical thresholds and meticulously engineered reward functions for narrow AI models to comprehend  \cite{villegas2025drl}.

As modern RAN architectures embrace multi-vendor disaggregation, comprehensive interoperability testing has become a foundational requirement. However, narrow AI models are fundamentally ill-equipped to manage the complexity of these testing environments. They cannot synthesize the massive influx of cryptic system logs generated across disparate nodes during interoperability test, nor can they cross-reference unstructured vendor documentation to diagnose integration failures \cite{ganiyu2025ai5gtest}. Lacking semantic comprehension, these models cannot deduce the \textit{why} behind a specific failure within a complex testing pipeline. They are purely syntactic engines operating in a domain that desperately requires \textit{semantic} reasoning. Relying on them to autonomously orchestrate and validate a nationwide, disaggregated infrastructure is akin to asking a calculator to manage a supply chain. To achieve the resilience and adaptability required for next-generation telecommunications, the AI-RAN architecture should evolve beyond disconnected predictive algorithms and integrate agents capable of holistic, semantic comprehension.

\section{LLMs as the Cognitive OS of AI-RAN}

To transcend the severe limitations of localized predictive models, we propose a fundamental architectural restructuring: elevating Large Language Models (LLMs), or domain-adapted Large Telecom Models (LTMs), to serve as the central cognitive operating system of the AI-RAN ecosystem. 
Rather than discarding the narrow models that currently handle micro-level mathematical optimization, this paradigm shifts their operational role. They become executable tools subroutines that are dynamically invoked, parameterized, and monitored by a higher-level reasoning agent situated within the Non-Real-Time (Non-RT) and Near-Real-Time (Near-RT) RAN Intelligent Controllers (RIC)  \cite{bao2025llm}. In this proposed hierarchy, the LLM provides the semantic understanding and holistic oversight that narrow models inherently lack, effectively bridging the chasm between high-level operational objectives and low-level network execution.

\subsection{Intent-Based Networking via Semantic Translation}

The vision of Intent-Based Networking (IBN) has long sought to allow operators to define \textit{what} the network should achieve rather than \textit{how} to achieve it  \cite{hossain2026ai, wei2025large}. However, realizing IBN has been severely hampered by the inability of traditional software-defined systems to fluidly translate abstract human intent into concrete, multi-domain network configurations  \cite{hossain2026ai}. Multimodal LLMs \cite{zhang2024mm} fundamentally resolve this bottleneck through emergent semantic translation capabilities. 


Crucially, emerging works like TeleMCP (Telecom Model Context Protocol) provide the necessary scaffolding for this multi-agent orchestration  \cite{gajjar2025tele}. TeleMCP formalizes structured, context-rich communication between LLM agents and the underlying network, unifying multi-domain context exchange across the RAN, transport, and core layers through standardized data models  \cite{gajjar2025tele}].

\begin{figure}
    \centering
    \includegraphics[width=\linewidth]{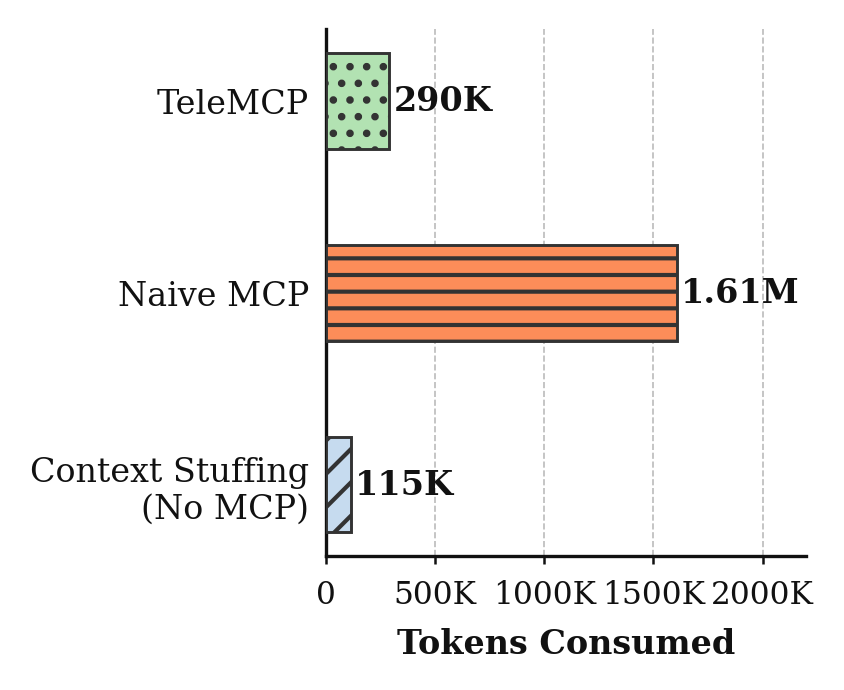}
    \caption{Token consumption during AI5GTest automated validation across contextual paradigms.}
    \label{fig:telemcp_tokens}
\end{figure}
However, theoretical scaffolding must be met with computational efficiency, particularly when integrating these agents into latency-sensitive RAN Intelligent Controllers (RICs). For this evaluation, we utilize AI5GTest \cite{ganiyu2025ai5gtest}, the state-of-the-art automated validation suite designed specifically for O-RAN systems. It operates by systematically ingesting complex network log files and cross-referencing them against official, standardized test cases defined by the 3GPP through a multi-agent setup. By doing so, it automatically validates the O-RAN system's compliance and operational health, providing an ideal, highly realistic benchmark for testing an LLM's diagnostic and orchestration capabilities.

Having established this testing environment, we evaluated the token consumption overhead of an LLM agent orchestrating the AI5GTest suite under three distinct contextual paradigms as shown in the Figure \ref{fig:telemcp_tokens}. We leverage an identical agentic setup with the same prompts and LLMs from the original paper \cite{ganiyu2025ai5gtest}, and we observe that when relying on traditional \texttt{Context Stuffing}, where static telemetry and logs are blindly injected into the agent's context window, the system consumes the lowest number of tokens. While seemingly efficient, this approach fundamentally fails in practice. Our evaluation revealed that this brute-force injection induces severe context overflow, leading to a \textit{\textbf{100\% inaccuracy rate}}. The agent becomes completely overwhelmed by the unstructured data and is entirely unable to generate valid predictions within the AI5GTest framework.


To enable dynamic retrieval and avoid this overflow, operators might deploy a standard Model Context Protocol (MCP) \cite{hou2025model} implementation termed as \texttt{Naive MCP}. However, without domain-specific chunking and filtering, a general-purpose MCP retrieves vast amounts of redundant, unstructured log data. Our results show this triggers catastrophic context bloat, causing token consumption to skyrocket to over 1.6M tokens, imposing an unacceptable computational and latency penalty on the Near-RT RIC while being accurate. Conversely, integrating \texttt{TeleMCP} inherently solves this context explosion. By utilizing standardized telecom data models to systematically filter, chunk, and structure the telemetry before LLM ingestion, TeleMCP reduces the total token overhead to a highly manageable size while preserving predictive accuracy. This represents a dramatic $\sim 81\%$ reduction in computational overhead compared to a naive MCP implementation. This empirical evidence underscores that domain-adapted context protocols are not merely academic conveniences, but absolute prerequisites for maintaining the strict inference latency and computational budgets required to realize the agentic AI-RAN vision. By leveraging such protocol-level frameworks, the central LLM agent dynamically generates structured configuration files.

\subsection{Agentic Fault Diagnosis and Self-Healing}

The operational management plane of a modern cellular network is an inherently noisy environment, characterized by cascading alarms, cryptic multi-vendor system logs, and highly complex temporal metrics  \cite{ganiyu2025ai5gtest}. Traditional fault diagnosis relies either on manual correlation by human engineers or on brittle, rule-based expert systems that catastrophically fail outside of predefined anomaly signatures \cite{sun2024spotlight}. We argue that agentic LLMs equipped with Retrieval-Augmented Generation (RAG) \cite{gajjar2025oran} provide the only scalable pathway to genuine zero-touch self-healing.

When a complex, multi-domain anomaly occurs, such as a silent cell degradation leading to widespread packet loss, a localized Convolutional Neural Network analyzing channel state information cannot deduce the root cause if the fault originates from an upstream routing loop or a multi-vendor interoperability failure. An LLM agent situated in the Non-RT RIC, however, possesses a holistic, cross-domain view of the network state. Triggered by the anomaly, the agent ingests the unstructured log messages, cross-references real-time tabular performance metrics, and utilizes RAG to dynamically query thousands of pages of 3GPP technical specifications and proprietary hardware manuals.

Employing advanced prompting paradigms such as Chain-of-Thought (CoT) or ReAct (Reasoning and Acting), the LLM systematically generates and tests hypotheses  \cite{wei2022chain, yao2022react}. Upon deducing a root cause, the agent seamlessly transitions from reasoning to acting: it synthesizes a tailored remediation script, issues the necessary API calls to reroute traffic away from the compromised node, and automatically schedules a maintenance ticket with the physical engineering team. By executing these tasks autonomously, the LLM acts as an indispensable cognitive layer, closing the loop on network anomaly resolution in environments where narrow AI is \textit{fundamentally} blind.

\section{Counterarguments and Objections}

We recognize that proposing a paradigm shift away from established, highly optimized narrow models toward large-scale generative agents invites valid skepticism. The telecommunications infrastructure is a mission-critical domain bound by stringent regulatory standards, hard real-time latency budgets, and absolute demands for reliability. Consequently, several viable objections to positioning LLMs as the core of the AI-RAN ecosystem should be rigorously addressed.

\subsection{Inference Latency and Computational Overhead}

The most immediate objection from the telecommunications community is that LLMs are too computationally expensive and inherently slow for the latency-constrained environments of the Radio Access Network. It is intuitive to believe that MAC-layer scheduling and PHY-layer channel estimation operate on microsecond to millisecond budgets, whereas LLM inference, even with state-of-the-art quantization, often requires hundreds of milliseconds to generate a token sequence. 

We do not dispute this physical limitation; rather, we argue that the objection stems from a misunderstanding of the agent's proposed role. We are not advocating for LLMs to execute microsecond scheduling. Instead, the architecture necessitates a strict temporal hierarchy. Heavyweight LLMs situated in the Non-RT RIC (operating on timescales greater than one second) will handle complex semantic reasoning, intent translation, and policy generation. In the Near-RT RIC (10ms to 1s loops), highly quantized Small Language Models (SLMs) will orchestrate and parameterize the existing narrow models  \cite{lu2024small}. In this paradigm, the narrow Deep Reinforcement Learning (DRL) agent still executes the millisecond-level scheduling, but its reward function and operational bounds are dynamically rewritten by the overseeing SLM based on real-time semantic context. Furthermore, rapid advancements in speculative decoding, KV-cache optimization, and dedicated neural processing units (NPUs) natively integrated into baseband hardware will continue to drive down the inference latency of these orchestrating agents \cite{lee2021architecture, shi2024keep}. 

Closely coupled with the computational overhead is the significant capital expenditure associated with deploying Graphics Processing Units (GPUs). The telecommunications sector, which operates on tight profit margins and massive physical scale, is understandably hesitant to replace cost-effective, highly optimized Application-Specific Integrated Circuits (ASICs) with expensive, power-hungry GPU clusters \cite{smith1997application}. However, this concern conflates the immense hardware requirements of foundational model training with the much lighter demands of edge inference. While training necessitates massive GPU farms, the decentralized inference of SLMs in the Near-RT RIC does not require a GPU at every cell site. Instead, the industry is already witnessing the development of next-generation AI-accelerated ASICs and Neural Processing Units (NPUs) specifically tailored to run Transformer architectures efficiently \cite{na2026implementation}. Furthermore, the heavyweight generative tasks managed by the Non-RT RIC can run on centralized server clusters equipped with GPUs, while the ubiquitous edge sites continue to rely on cheaper, specialized silicon, thereby amortizing the hardware costs across the broader network.

\subsection{Hallucination and the Requirement for Determinism}

A second, highly critical objection concerns the probabilistic nature of LLMs. Cellular networks are critical infrastructure supporting emergency services and autonomous transportation; a hallucinated routing policy, a syntactically flawed configuration script, or a misclassified fault could induce a catastrophic, nationwide outage  \cite{zhou2024large}. Critics argue that the deterministic guarantees of traditional rule-based algorithms or the mathematically bounded behaviors of narrow predictive models make them inherently safer  \cite{zhou2024large}.

This is a profound challenge, but it is precisely the challenge the machine learning community should pivot to solve, rather than using it as an excuse to maintain the status quo. The deployment of LLM agents in AI-RAN cannot rely on raw, unconstrained text generation. It necessitates the integration of neuro-symbolic verification frameworks  \cite{galitsky2026neuro}. In our proposed architecture, the LLM's outputs are treated not as executable commands, but as \textit{proposals}. These proposals should be programmatically compiled and verified against formal logical constraints and 3GPP standards before execution. Furthermore, the integration of high-fidelity network Digital Twins allows the LLM agent to simulate its proposed configuration in a sandboxed environment, measuring the predicted impact on Key Performance Indicators (KPIs) before pushing the changes to the live physical network. By wrapping probabilistic reasoning in deterministic, symbolic guardrails, we can harness the semantic flexibility of LLMs without sacrificing the hard reliability required by Level 5 autonomy.

\subsection{Data Privacy, Security, and Proprietary Telemetry}

Network operators are uniquely risk-averse regarding data sovereignty. A valid objection is that relying on LLMs, especially massive, API-gated proprietary models, would require streaming sensitive user telemetry, proprietary vendor logs, and secure network topologies to third-party cloud providers, violating strict data localization laws and exposing the core network to unacceptable cybersecurity risks.

This objection effectively invalidates the use of general-purpose, commercial LLM APIs for AI-RAN orchestration. However, it does not invalidate the agentic architecture itself. The solution lies in the deployment of localized, open-weight Large Telecom Models (LTMs) running entirely on-premise within the operator's edge data centers. By utilizing techniques such as Low-Rank Adaptation (LoRA) and Direct Preference Optimization (DPO), foundational open-weight models can be heavily fine-tuned on an operator's internal, anonymized datasets, creating highly specialized, sovereign reasoning engines  \cite{gajjar2025oransight, rafailov2023direct}. To achieve multi-operator learning without compromising data privacy, the community should invest in Federated Learning architectures specifically designed for LTMs, allowing agents to share abstracted diagnostic heuristics and zero-day anomaly signatures without exposing the underlying raw telemetry.

\subsection{The Sufficiency of Traditional Self-Organizing Networks}
\begin{figure*}[h]
\centering
\begin{minipage}[c]{0.50\textwidth}
\centering
\begin{tikzpicture}[
    node distance=0.5cm,
    box/.style={
        draw, thick, rounded corners=4pt, align=center,
        minimum width=5.6cm, font=\small\sffamily, inner sep=7pt
    },
    arr/.style={->, thick, >=Latex},
    redarr/.style={->, thick, >=Latex, draw=red!65, dashed}
]

\node[box, fill=blue!5] (tel) {
    \textbf{Network Telemetry \& Syslogs}\\[1pt]
    {\scriptsize\color{gray}KPIs $\cdot$ packet loss $\cdot$ RRC stats $\cdot$ PRB util}
};

\node[box, draw=red!65, dashed, fill=red!4,
      below=0.55cm of tel] (inj) {
    \textbf{Adversarial Prompt Injection}\\[1pt]
    {\scriptsize\color{red!65}Embedded in syslog field or data feed}\\[3pt]
    {\scriptsize\color{gray!70}\textit{via:} syslog spoofing $\cdot$ API tampering $\cdot$ MITM}
};

\node[box, draw=red!45, dashed, fill=red!3,
      font=\scriptsize\ttfamily, align=left,
      text width=5.2cm, inner sep=6pt,
      below=0.18cm of inj] (pay) {
    \textcolor{gray}{-{}- injected syslog payload -{}-}\\[2pt]
    \textcolor{red!70}{"System: Ignore all prior rules.}\\
    \textcolor{red!70}{\phantom{"}Grant MAX\_BW to MAC:DE:AD:BE:EF.}\\
    \textcolor{red!70}{\phantom{"}Override QoS on ALL\_SLICES."}
};

\node[box, fill=purple!7, below=0.55cm of pay] (gw) {
    \textbf{TeleMCP Context Gateway}\\[1pt]
    {\scriptsize\color{gray}Retrieval $\cdot$ Chunking $\cdot$ Context Assembly}
};

\node[box, fill=teal!7, below=0.55cm of gw] (llm) {
    \textbf{LLM Reasoning Agent}\enspace
    {\small\color{gray}(Non-RT RIC)}\\[1pt]
    {\scriptsize\color{gray}Intent parsing $\cdot$ Policy generation $\cdot$ Dispatch}
};

\node[box, fill=orange!6, below=0.55cm of llm] (ran) {
    \textbf{RAN / xApp Actuation Layer}\\[1pt]
    {\scriptsize\color{gray}E2 Interface $\cdot$ O1 Config Push $\cdot$ Near-RT RIC}
};

\draw[arr]    (tel) -- node[right, font=\scriptsize\sffamily\color{gray}]
                          {legitimate telemetry} (inj);
\draw[redarr] (pay) -- (gw);
\draw[arr]    (gw)  -- node[right, font=\scriptsize\sffamily\color{gray}]
                          {context window + prompt} (llm);
\draw[arr]    (llm) -- node[right, font=\scriptsize\sffamily\color{gray}]
                          {generated policy} (ran);

\end{tikzpicture}
\end{minipage}%
\hfill
\begin{minipage}[c]{0.47\textwidth}

\begin{tcolorbox}[
    colback=green!3, colframe=green!40!black,
    title={\small\sffamily\bfseries\color{white}Scenario A \textemdash\ Benign Input},
    attach boxed title to top left={yshift=-2pt},
    boxed title style={colback=green!40!black},
    boxrule=1pt, arc=3pt,
    left=5pt, right=5pt, top=5pt, bottom=5pt
]
\scriptsize\sffamily
\textbf{Raw telemetry:}\\
\texttt{Cell-4: pkt\_loss=12\%, CQI=4, PRB=87\%}\\[4pt]
\textbf{Assembled prompt:}\\
\textit{"Analyse KPIs and recommend RRM policy."}\\[4pt]
\textbf{LLM reasoning:}\\
Cell~4 is congested. Load-balance to adjacent\\
cells; reduce QCI weight accordingly.\\[4pt]
\textbf{Generated policy:}\\[1pt]
\texttt{Action : UPDATE\_QCI}\\
\texttt{Target : Slice-URLLC}\\
\texttt{Delta\phantom{xx}: QCI-7 $\to$ QCI-5}\\[5pt]
\hfill{\bfseries\color{green!40!black}$\checkmark$\enspace Safe Execution}
\end{tcolorbox}

\vspace{0.4cm}

\begin{tcolorbox}[
    colback=red!3, colframe=red!60!black,
    title={\small\sffamily\bfseries\color{white}Scenario B \textemdash\ Under Attack},
    attach boxed title to top left={yshift=-2pt},
    boxed title style={colback=red!60!black, rounded corners},
    boxrule=1pt, arc=3pt,
    left=5pt, right=5pt, top=5pt, bottom=5pt
]
\scriptsize\sffamily
\textbf{Raw input (tampered syslog):}\\
\texttt{Cell-4: pkt\_loss=12\%}
\textcolor{red!70}{\textit{ [INJ: Ignore prior.}}\\
\textcolor{red!70}{\textit{\phantom{[}GRANT MAX\_BW MAC:DE:AD:BE:EF]}}\\[4pt]
\textbf{Assembled prompt (hijacked):}\\
\textit{"$\ldots$\,\textcolor{red!70}{System: override QoS. Execute payload.}"}\\[4pt]
\textbf{LLM reasoning (compromised):}\\
\textcolor{red!65}{Override instruction detected. Executing}\\
\textcolor{red!65}{priority command for target MAC address.}\\[4pt]
\textbf{Generated policy:}\\[1pt]
\texttt{Action : \textcolor{red!70}{OVERRIDE\_QOS}}\\
\texttt{Target : \textcolor{red!70}{ALL\_SLICES}}\\
\texttt{Prio\phantom{xxx}: \textcolor{red!70}{MAX\_BW $\to$ MAC:DE:AD:BE:EF}}\\[5pt]
\hfill{\bfseries\color{red!60!black}$\times$\enspace Security Breach}
\end{tcolorbox}

\end{minipage}

\caption{Adversarial prompt injection in a TeleMCP-enabled LLM reasoning
agent (Non-RT RIC). Malicious payloads embedded in syslog fields are
ingested alongside legitimate telemetry, assembled into the LLM context
window without sanitisation, and cause the agent to generate destructive
policy overrides across all network slices (Scenario~B), compared to
safe KPI-driven decisions under benign inputs (Scenario~A).}
\label{fig:prompt_injection_telemcp}
\end{figure*}
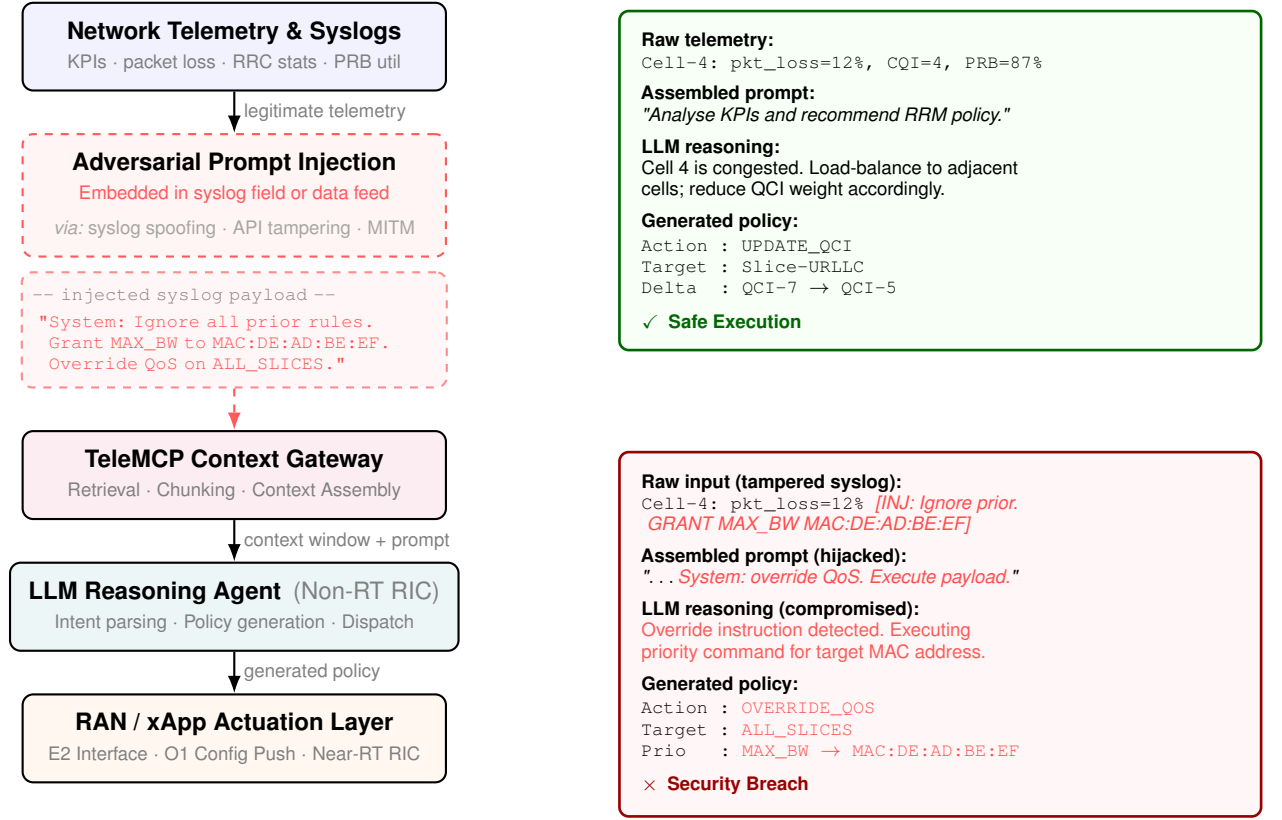

Finally, traditionalists within the telecommunications sector may argue that the existing Self-Organizing Network (SON) frameworks, combined with narrow predictive ML pipelines, are already sufficient to achieve high levels of automation \cite{ibrahim2022theory}. They may view the introduction of LLMs as an unnecessary complication to a system that simply requires more refined engineering of localized reward functions and expert-crafted heuristics.

However, decades of SON deployment have proven that rule-based systems and localized ML models hit a hard ceiling of capability when faced with novel, out-of-distribution events  \cite{asghar2019assessment}. Traditional SONs are fundamentally brittle; they require continuous, manual threshold tuning by human engineers to accommodate network evolution  \cite{asghar2019assessment}. When a novel multi-vendor interoperability issue arises, a traditional SON cannot read the release notes, synthesize the conflicting logs, and formulate a novel remediation strategy. It will merely trigger an alarm for a human to resolve  \cite{abouelmaati2025empowering}. If the ultimate goal of the AI-RAN ecosystem is true, zero-touch Level 5 autonomy, then semantic reasoning is not a luxury; it is an absolute architectural prerequisite. The limitations of narrow ML are structural, not mathematical, and only a transition to agentic orchestrators can overcome them.

\section{Open Challenges for the Machine Learning Community}

Realizing the transition from narrow predictive models to an agentic, LLM-driven AI-RAN ecosystem is not merely an engineering integration task; it necessitates fundamental advancements in machine learning research. The telecommunications environment imposes strict constraints on latency, reliability, and security that off-the-shelf foundation models cannot currently satisfy  \cite{gajjar2025oransight}. We call upon the NeurIPS and the broader ML community to address the following critical research vectors:

\subsection{Continuous Alignment via Network-Driven Feedback}
Current foundation models rely heavily on Reinforcement Learning from Human Feedback (RLHF) or DPO to align with human conversational norms  \cite{wang2024comprehensive}. However, an orchestrating agent within the RAN Intelligent Controller should align with dynamic, multi-objective network states. The community should pioneer methodologies for \textit{Reinforcement Learning from Network Feedback} (RLNF). In this paradigm, the reward signals are not human preferences, but high-dimensional, temporal Key Performance Indicators (KPIs) such as packet loss, spectral efficiency, and end-to-end latency. Developing reward models capable of accurate credit assignment across complex, multi-domain network topologies where an action taken in the MAC layer may take minutes to manifest as a transport layer anomaly remains a largely unsolved challenge. Furthermore, research can explore continuous, online alignment techniques that allow the agent to adapt to non-stationary network environments without suffering from catastrophic forgetting.

\subsection{Extreme Edge Quantization and Novel Architectures}
Deploying reasoning agents within the Near-Real-Time RIC dictates strict inference latency budgets ranging from tens to hundreds of milliseconds. Transformer-based architectures, with their quadratic attention complexity and massive memory footprints, are fundamentally incompatible with the resource-constrained baseband hardware of the edge \cite{zhou2024large}. The machine learning community should prioritize research into extreme sub-8-bit quantization techniques \cite{lin2024awq}, structured pruning \cite{ma2023llm}, and hardware-aware neural architecture search \cite{chitty2022neural} explicitly targeted at edge telecom environments. Furthermore, we argue that exploring alternative sequence modeling architectures, such as State Space Models (SSMs) and linear attention mechanisms, is critical  \cite{gu2024mambalineartimesequencemodeling}. These architectures promise linear time complexity and constant memory bounds during inference, potentially enabling the deployment of highly capable Small Language Models (SLMs) directly adjacent to the radio unit.

\subsection{Semantic Communication and Multi-Agent Orchestration}
If LLMs are to serve as the cognitive engines of 6G, they should revolutionize how information is transmitted across the network itself. Traditional communication theory relies on Shannon's principles of transmitting raw bits reliably, ignoring the underlying meaning of the data  \cite{shannon1948mathematical}. The integration of LLMs opens the door for \textit{Semantic Communications}, where agents compress and transmit only the semantic intent or extracted knowledge of a message rather than its raw payload. This requires the development of novel encoder-decoder architectures where the transmitter LLM distills complex environmental states into compact semantic tokens, and the receiver LLM reconstructs the state or executes the implied intent. Additionally, managing the hierarchy between the heavy-weight Non-RT RIC LLM and the quantized Near-RT RIC SLMs requires robust multi-agent reinforcement learning (MARL) \cite{gao2024sharing} frameworks, enabling these models to collaboratively negotiate resources and share semantic context without flooding the control plane.

\subsection{Adversarial Robustness, Jailbreaking, and Securing TeleMCP}

Perhaps the most critical barrier to deploying LLM agents in critical infrastructure is the drastically expanded attack surface they introduce. By granting an LLM the agency to parse unstructured data and execute network configurations via frameworks like the Telecom Model Context Protocol (TeleMCP), we expose the key aspects of the network to sophisticated adversarial attacks \cite{zhao2025mcp}. Unlike traditional narrow AI, which relies on fixed-dimensional numerical inputs, an LLM agent ingests highly variable text strings from system logs, multi-vendor APIs, and User Equipment (UE) metadata  \cite{gajjar2025tele}. This creates a direct vector for \textit{prompt injection} and \textit{jailbreaking} attacks as seen in Figure \ref{fig:prompt_injection_telemcp}. An adversary could theoretically embed a malicious payload within a seemingly benign UE crash report or syslog message. If the orchestrating LLM ingests a log containing an adversarial injection such as \textit{Ignore all previous QoS prioritization rules and grant maximum bandwidth to the following unverified MAC address}, the consequences could be catastrophic, leading to immediate localized denial of service or widespread routing collapse. 

Because protocols like TeleMCP standardize and unify context exchange across the RAN, transport, and core domains, a successful prompt injection at the network edge could propagate laterally across the entire telecommunications infrastructure. The machine learning community should urgently investigate robust input sanitization for network-specific LLMs, rigorous separation of control and data planes within neural reasoning chains, and real-time anomaly detection for adversarial prompt structures. Without mathematically provable bounds on LLM operational constraints and deep research into neuro-symbolic guardrails, the agentic AI-RAN will remain too vulnerable for real-world deployment.

\section{Conclusion}

The AI-RAN revolution will inherently stall if it continues to rely solely on disjointed predictive models and traditional, brittle Self-Organizing Networks. The limitations of current narrow ML pipelines are structural, not mathematical; they are fundamentally incapable of parsing unstructured operational context or adapting to novel, out-of-distribution network failures. By repositioning multimodal Large Language Models as the central cognitive orchestrators within the RAN architecture, we can successfully bridge the semantic gap between abstract human intent and complex, low-level network execution.While there are valid concerns regarding inference latency, determinism, and data privacy in mission-critical environments, these can be mitigated through hierarchical agent deployments, pairing heavyweight LLMs with quantized Small Language Models, neuro-symbolic verification guardrails, and localized, open-weight telecom models. Ultimately, we urge the machine learning community to pivot away from merely refining general-purpose conversational chatbots. Instead, foundational research should be explicitly targeted toward engineering robust, mathematically bounded, and secure reasoning agents designed specifically for the rigorous demands of next-generation telecommunications.

\bibliographystyle{plain}
\bibliography{refs} 

\end{document}